\documentclass[fleqn,usenatbib]{mnras}

\usepackage[final]{changes}
\usepackage{graphicx}
\usepackage{amsmath}
\usepackage{amssymb}
\usepackage{booktabs}
\usepackage{latexsym}
\usepackage{multirow}
\usepackage{hyperref}
\usepackage{float}
	
\hypersetup{
    colorlinks,
    citecolor=blue,
    filecolor=blue,
    linkcolor=blue,
    urlcolor=blue
}
\colorlet{Changes@Color}{orange}

\begin{document}

\title{Planetary Magnetic Field Control of Ion Escape from Weakly Magnetized Planets}

\author[H. Egan]{Hilary Egan$^{1}$\thanks{Contact e-mail: \href{mailto:Hilary.Egan@colorado.edu}{Hilary.Egan@colorado.edu}},
 Riku Jarvinen$^{2,3}$, Yingjuan Ma $^{4}$,  David Brain$^{1}$
\
\\
$^{1}$ Department of Astrophysical and Planetary Sciences, University of
   Colorado, Boulder, CO 80309, USA ,\\
$^{2}$ Department of Electronics and Nanoengineering, School of Electrical Engineering, Aalto University, Espoo, Finland,\\
$^{3}$ Finnish Meteorological Institute, Helsinki, Finland\\
$^{4}$
Department of Earth Planetary and Space Sciences, University of California, Los Angeles, Los Angeles, CA, USA}

\maketitle

\begin{abstract}
    Intrinsic magnetic fields have long been thought to shield planets from atmospheric erosion via stellar winds; however, the influence of the plasma environment on atmospheric escape is complex. Here we study the influence of a weak intrinsic dipolar planetary magnetic field on the plasma environment and subsequent ion escape from a Mars sized planet in a global three-dimensional hybrid simulation. We find that increasing the strength of a planet's magnetic field enhances ion escape until the magnetic dipole's standoff distance reaches the induced magnetosphere boundary. After this point increasing the planetary magnetic field begins to inhibit ion escape. This reflects a balance between shielding of the southern hemisphere from ``misaligned" ion pickup forces and trapping of escaping ions by an equatorial plasmasphere. Thus, the planetary magnetic field associated with the peak ion escape rate is critically dependent on the stellar wind pressure. Where possible we have fit power laws for the variation of fundamental parameters  (escape rate, escape power, polar cap opening angle and effective interaction area)  with magnetic field, and assessed upper and lower limits for the relationships.
\end{abstract}

\section{Introduction}
\label{sec:intro}

Atmospheric escape is capable of shaping a planet’s atmospheric composition and total mass, and thus the planet's long-term habitability. Loss to space of atmospheric particles is thought to have played a key role in the atmospheric evolution of both Mars \citep[e.g.][]{1977Sci...198..453A, 2001Natur.412..237J, 2018Icar..315..146J} and Venus \citep[e.g.][]{Dayhoff556, KULIKOV20061425}. This evolution may have prevented these planets from being currently habitable like the Earth, despite their likely similarities in initial formation. As we begin to discover an abundance of exoplanets, it is important to understand the processes that shape their evolution, including atmospheric escape. This will be particularly interesting for potentially habitable terrestrial planets, such as Proxima-b \citep{2016Natur.536..437A, 2018AsBio..18..133M} and those in the Trappist-1 system \citep{2017Natur.542..456G, 2018PNAS..115..260D}. 

Hydrogen atmospheres accreted early in a planet's life are thought to escape due to the thermal energy of the particles, either slowly through Jean’s escape or quickly due to a thermal wind \citep{1973JAtS...30.1481H}. In both of these processes a significant portion of the particle velocity distribution exceeds escape velocity due to their thermal motion. This thermal energy can be gained from photo-heating \citep[e.g.][]{2004Icar..170..167Y}, giant impacts \citep{2015ApJ...812..164L}, or core-accretion \citep{2016ApJ...825...29G}. Thermally-driven atmospheric escape is thought to lead to the gap between super-earths and mini-neptunes \citep{2017ApJ...847...29O, 2018MNRAS.476..759G, 2018arXiv180906810B}.

Although thermal loss is an important process for dictating the evolution of the primary atmosphere or hydrogen nebula of a planet, secondary atmospheres created by planetary out-gassing are composed of gravitationally bound heavier species and have more difficulty escaping thermally after the primary atmosphere has been lost. Additional energy sources such as photo-chemical production  \citep{1977JGR....82.4379M, 1982P&SS...30..773H}, sputtering \citep{1979JGR....84.8436H, 1982P&SS...30..773H}, charge exchange \citep{1977JGR....82....1C}, or ion escape are necessary to drive significant escape of secondary atmospheres over the course of a planet’s lifetime \citep{1982P&SS...30..773H}. Ion escape encompasses the collection of escape channels where escape occurs through loss of ionized species, including ion pickup \citep[e.g.][]{1991JGR....96.5457L}, ion bulk escape \citep[e.g.][]{2010GeoRL..3714108B}, and the polar wind \citep[e.g.][]{1968P&SS...16.1019B, 2007JASTP..69.1936Y}. Both the energization of the particles through these processes and their eventual escape trajectories (or lack thereof) are critically affected by the dynamic plasma environment created by the interaction of the planet and the stellar wind.

The strength of a planet's magnetic field is a critical factor in shaping a planetary plasma environment. In the solar system the only clearly habitable planet, Earth, has a strong intrinsic magnetic field driven by an internal dynamo, while Venus and Mars do not. This has led many to speculate that an intrinsic magnetic field is a critical component required to shield the planet from atmospheric erosion due to the solar wind \citep{DRISCOLL20131447, Lundin2007, 2016ApJ...820L..15D,Driscoll2018}, we call this the magnetic umbrella hypothesis. However, present measurements of escape rates suggest that all three planets are losing their atmosphere at similar rates \citep{2005JGRA..110.3221S}, and the magnetic umbrella hypothesis has been called into question \citep{kasting2012find, 2013cctp.book..487B}. Understanding the complex interplay between the planetary plasma environment, solar wind driving, and ion loss is thus of critical importance. 

Some investigation into the effects of an intrinsic magnetic field on ion loss has been made using global plasma models. \citet{ 2017ApJ...837L..26D} compared escape rates from Proxima-b using a multi-species MHD model with a magnetized and an unmagnetized case and found larger escape rates in the unmagnetized case, while \citet{2017ApJ...844L..13G} found using a polar wind outflow model that the higher ionizing radiation levels of EUV radiation fluxed of M dwarfs compared to Earth may increase the effectiveness of the polar wind by orders of magnitude. Global hybrid modeling work from \citet{2012EP&S...64..149K} looked at ion escape from a Mars type planet with a global dipole of 0, 10, 30, and 60 nT, and found maximum ion escape from the 10 nT case. \citet{gunell_2018} combined empirical measurements at Earth, Mars, and Venus with semi-analytic models to analyze the influence of planetary magnetic field over a large range of magnetic field values, finding multiple peaks in escape rate over the range that varied in strength and value by planet. \citet{doi:10.1029/2018GL079972} found an increased escape rate for a weakly magnetized Mars (100~nT) compared to an unmagnetized Mars in a multi-species MHD model. In combination these results suggest that there is much more to learn about about ion escape as it relates to planetary magnetic fields.

Weakly magnetized planets are an especially interesting regime as they represent the transition from unmagnetized planets with induced magnetospheres to magnetized planets with intrinsic magnetospheres. In the weak-field limit the scale size of the stellar wind interaction region is not dramatically changed. However, a global planetary magnetic field is typically approximated to the first order as a magnetic dipole centered at the planet, this introduces a new axis of symmetry. Furthermore, some studies indicate that planets around M-dwarfs may be weakly magnetized \citep{2010ApJ...718..596G}. Analysis of the Kepler data has shown that planetary systems are common around M-dwarfs \citep{2013ApJ...767L...8K}, and these systems also show the best promise of observing exoplanet atmospheres \citep{2016PhR...663....1S}. Thus it is important to understand what effect weak magnetic fields will have on the subsequent ion escape.

Here we study the influence of weak planetary magnetic fields on the plasma environment of a planet and subsequent ion loss through a series of hybrid plasma simulations. Section \ref{sec:methods} describes our methods, Section \ref{sec:plasma_env} details influence of the magnetic field on the planetary plasma environment, Section \ref{sec:ion_esc} describes the corresponding influence on ion escape, Section \ref{sec:discussion} contains a discussion of the results, and Section \ref{sec:conclusions} summarizes the conclusions.
\section{Methods}
\label{sec:methods}

To analyze the effects of planetary magnetic field strength on ion escape we have run seven simulations of the solar wind interaction with increasingly magnetized planets. These simulations were performed using the hybrid plasma model RHybrid \citep{2018JGRA..123.1678J}. In a hybrid model the ions are treated as macroparticle clouds while the electrons are treated as a charge-neutralizing fluid. By treating ions as macroparticles ion kinetic effects associated with finite ion gyroradii are included, while treating the electrons as a fluid allows simulating a much larger volume than a corresponding fully kinetic approach.

Each ion macroparticle represents a group of ions that have the same velocity ($v_i$), central position ($x_i$), charge ($q_i$), and mass ($m_i$). Each cloud has the same shape and size as a grid cell and clouds move inside the grid cells obeying the Lorentz force such that

\begin{equation}
m_i \frac{d\vec{v_i}}{dt} = q_i(\vec{E}+\vec{v_i}\times\vec{B})\;,
\end{equation}

where $\vec{E}$ and $\vec{B}$ are the electric and magnetic fields respectively. The electron charge density then follows from the quasi-neutral assumption with all ion species accumulated in the grid and total new charge density assumed to be zero in each cell.

Ampere’s law is used to calculate the current density from the magnetic field

 \begin{equation}
\vec{J} = \nabla\times\vec{B}/\mu_0 \;,
\end{equation} and then the electric field is found using Ohm's law

\begin{equation}
\vec{E} = -\vec{U_e}\times \vec{B}+\eta\vec{J}\;,
\end{equation}

where the $\eta$ is the explicit resistivity profile, and $U_e$ is the electron fluid velocity. The value of $\eta$ was chosen to be $\eta = 0.02\mu_0\Delta x^2/\Delta t$, such that the magnetic diffusion time scale becomes  $\tau_D = \mu_0 L_B^2/\eta=50\Delta t$, for the magnetic length scale $L_B\simeq\Delta x$. This choice adds diffusion in the propagation of the magnetic field \citep{2008SSRv..139..143L}, stabilizing the numerical integration, while ensuring that the magnetic field diffuses on timescales longer than the timestep $\Delta t$ . Note that the resistivity is not explicitly included in the Lorentz force. Finally, the magnetic field is advanced using Faraday's Law
\begin{equation}
\frac{\partial \vec{B}}{\partial t} = -\nabla \times \vec{E}\;.
\end{equation}

\citet{2018JGRA..123.1678J} and references therein contain details of the numerical scheme. Highly parallelized RHybrid and its sequential predecessor code HYB have been used to study ion escape from Mars \citep{2002JGRA..107.1035K, 2010Icar..206..152K, JGRA:JGRA21894}, Venus \citep{2009AnGeo..27.4333J, 2013JGRA..118.4551J}, Mecury \citep{2003AnGeo..21.2133K}, Titan \citep{2007GeoRL..3424S09K}, and exoplanets around M-dwarfs \citep{egan_2019}.

The results presented in this paper are in the Planet Stellar Orbital (PSO) coordinate system, such that the planet is at the origin, the x direction points from the planet against the undisturbed upstream stellar wind velocity vector, the z direction is perpendicular to the orbital plane of the planet, and y is the completion of a right-handed coordinate system. The physical distances are shown in planetary radii ($R_p$) that correspond to 3390~km ($=1$ Martian radius).

\begin{table}[]
\centering
\caption{Upstream conditions used to drive the models. Here $u$ is the stellar wind velocity, $n_p$ is the stellar wind $H^+$ number density, $T_p$ is the stellar wind $H^+$ temperature, $B_{sw}$ is the interplanetary magnetic field (IMF), $P_{sw}$ is the dynamical pressure, and $v_{A}$ is the Alfven speed.}
\label{tab:sw_params}

\begin{tabular}{ll}\hline
 $u$ &  $[-350,0,0]\;\mathrm{km/s}$\\
 $n_p$ & 4.9 $\mathrm{cm}^{-3}$ \\
 $T_p$ & 59200 K \\
 $B_{sw}$ & $[-0.74, 5.46, -0.97]\;\mathrm{nT}$ \\
 $P_{sw}$ & 1.0 nPa\\
 $v_{A}$ & 55 km/s\\ \hline
\end{tabular}
\end{table}

The externally forced stellar wind and interplanetary magnetic field (IMF) was chosen to be the same as in \citet{egan_2018a} and is a case of a nominal solar wind interaction with Mars. The corresponding parameters are listed in Table \ref{tab:sw_params}. Note that the magnetic field is largely in the +y direction and thus the corresponding motional electric field is largely in the +z direction.

The ion production representative of the planetary ionosphere is implemented via Chapman profiles such that 

\begin{equation}
	q(\chi, z') = Q_0 \exp[1-z'-\sec(\chi)*e^{-z'}]\;,
\end{equation}

where $z'$ is the normalized height parameter given by $z' = (z-z_0)/H$, $\chi$ is the solar zenith angle, $z_0$ is the reference height, and $H=16$ km is the scale height. Such a profile assumes an isothermal atmosphere that is ionized by plane-parallel, monochromatic radiation in the EUV. This is a simplified method to inject planetary ionospheric ions in the simulation using a pre-defined production rate and spatial emission profile near the exobase and the inner boundary of the model; limitations arising from such a choice are discussed in Section \ref{sec:discussion}. The peak location of ion production occurs at an altitude of $z_0=300$~km, while the ion absorption lower boundary occurs at 150 km. The global emission rates $Q_{0}$ for each ion are picked to correspond to nominal values at Mars today ($Q_0(O_2^+)=2\times10^{25}$~s$^{-1}$, $Q_0(O^+)=1.4\times 10^{25}$~s$^{-1}$) \citep{2016P&SS..127....1J, 2018JGRA..123.1678J, egan_2018a, egan_2019}. 

The planetary magnetic field is enforced at the planet surface, and in the non-magnetized case is implemented as a super-conducting sphere. The range of planetary magnetic fields runs from unmagnetized to an equatorial surface dipole field of $B_p=150$~nT, which corresponds to approximately 1/400th of the Earth’s equatorial surface magnetic field. The magnetic dipole is aligned with the z axis such that equatorial dipole field point along $+z$.

Each simulation was run on a $240^3$ grid, with boundaries at $\pm 5 R_P$ in the Y and Z directions and $[-6,+4]$ in the X direction. This leads to a corresponding resolution of $\Delta x = 141$ km, which is comparable to an estimate of the ion gyroradius for $O^+$ ($\gamma = mv_\perp/qB\sim390$~km). Each simulation was run for 100,000 time steps with $\Delta t = 0.01 s$ (satisfying the Courant condition), corresponding to roughly 8 stellar wind crossing times through the simulation domain.
\section{Unmagnetized and Magnetized Planetary Paradigms}
\label{sec:theory}

For unmagnetized planets with some atmosphere, the conductivity of the ionosphere induces a magnetosphere which acts to stand off the stellar wind in the absence of a planetary magnetic field. This creates a transition region that includes the Magnetic Pileup Boundary (MPB) due to the pileup of the IMF as it drapes around the planet and the Induced Magnetosphere Boundary (IMB) marked by a transition from stellar wind to planetary plasma \citep{2004SSRv..111...33N}. The dynamics of ion escape are thus dominated by the plasma environment associated with the induced magnetosphere.

As the stellar wind directly interacts with the ionosphere at unmagnetized planet, the stellar wind motional electric field accelerates pickup ions into a plume \citep{CLOUTIER1974967, 1991JGR....96.5457L, doi:10.1002/2015GL065346}. Additionally, cold ions can flow out on the night-side of the planet in tail escape \citep{FRANZ201592, engwall2009}.

For magnetized planets the intrinsic magnetic field  acts to stand off the incoming stellar wind through a balance of the magnetic pressure and incoming stellar wind dynamic pressure. For a planetary magnetic field given by a dipole this pressure balance can be used to define the magnetic standoff distance Rs as follows

\begin{equation}
    \label{eq:standoff}
    R_s = R_P \left(\frac{P_B}{P_{sw}}\right)^{1/6} = R_P \left(\frac{2B_P^2}{\mu_0 \rho_{sw} u_{sw}^2}\right)^{1/6}  
\end{equation}

where the additional factor of 2 in the numerator comes from the compression of the magnetic field at the magnetopause \citep{2004pssp.book.....C}. Because the stellar wind has less direct access to the ionosphere, the dynamics of the escaping ions are dictated by the plasma environment associated with the intrinsic magnetosphere and the topology of the field lines. 

Magnetic topology is defined by the structure of magnetic field lines and their relationship to the surrounding environment. Open magnetic field lines have one end connected to the planet with the other connected directly to the stellar wind, while closed field lines have both foot-points connected to the planet. As a planet's magnetic field increases, the area of the ionosphere that is magnetically connected to the stellar wind via open field lines can be expected to decrease.

The locations of closed and open field lines can be estimated analytically for magnetized planets using the magnetic stand-off distance. Because the magnetic standoff distance is associated with the last closed field line, the latitude of the polar cap, or region of open field lines is thus given by

\begin{equation}
    \label{eq:theta_crit}
    \theta_c = \arcsin\left(\sqrt{\frac{R_p}{R_s}}\right)
\end{equation}

and the corresponding polar cap solid angle is then

\begin{equation}
    \label{eq:cap_angle}
    \Omega_c = 4\pi(1-\cos(\theta_c)) = 4\pi(1-\sqrt{1-R_p/R_s}).
\end{equation}

When the standoff distance becomes less than the planet radius the polar cap angle becomes saturated at $4\pi$, or twice the angle of an entire hemisphere.

\added{As ions are constrained to gyrate around magnetic field lines, magnetic topology has key implications for determining ion escape rates.} At strongly magnetized planets ions escape along open field lines at the poles in a process known as cusp escape \citep{doi:10.1029/JA090iA05p04099}. If the ions are accelerated due to the electron pressure gradients along the poles it is called the polar wind \citep{1968P&SS...16.1019B, 2007JASTP..69.1936Y}.

\added{The transition between magnetized and unmagnetized systems is a transition between these two paradigms; unmagnetized planets are dominated by induced magnetosphere dynamics and direct acceleration of planetary ions via the stellar wind motional electric field, while magnetized planets are dominated by their intrinsic magnetospheres and global magnetic topology. The following results sections explore this transition in plasma environments and the corresponding changes ion escape.}
\section{Plasma Environment}
\label{sec:plasma_env}

In this section we analyze the structure of the plasma environment and magnetic topology near the planet in our set of runs with an increasing planetary magnetic field strength. Even a moderate change in environment and topology associated with a weakly magnetized planet can affect the ion dynamics near the planet and thus the ion outflow.

\subsection{Tail Twisting}

The tail magnetic morphology is critically affected by both the IMF and the dipole structure. In an unmagnetized planet the stellar wind magnetic field drapes around the planet in a roughly symmetric fashion, creating a central current sheet \added{within the induced magnetosphere}. Observations at Mars have shown that the current sheet is very dynamic with local variations ascribed to both twisting and flapping motions \citep{doi:10.1002/2016JA023488,2018GeoRL..45.4559D,2019ApJ...871L..27C}, which may be associated with changes in response to upstream IMF or stellar wind pressure,  kink-like flapping due to waves propagating along the current sheet, or reconnection of crustal fields \citep{2015GeoRL..42.9087L}. For magnetized planets the \added{intrinsic} dipole field also creates a current sheet in the tail oriented perpendicularly to the dipole axis. This current sheet also experiences flapping motion \citep{doi:10.1029/JZ072i001p00131} and is strongly influenced by reconnection events \citep{2006JGRA..11111206N}.

\begin{figure*}
    \centering
    \includegraphics[width=\textwidth]{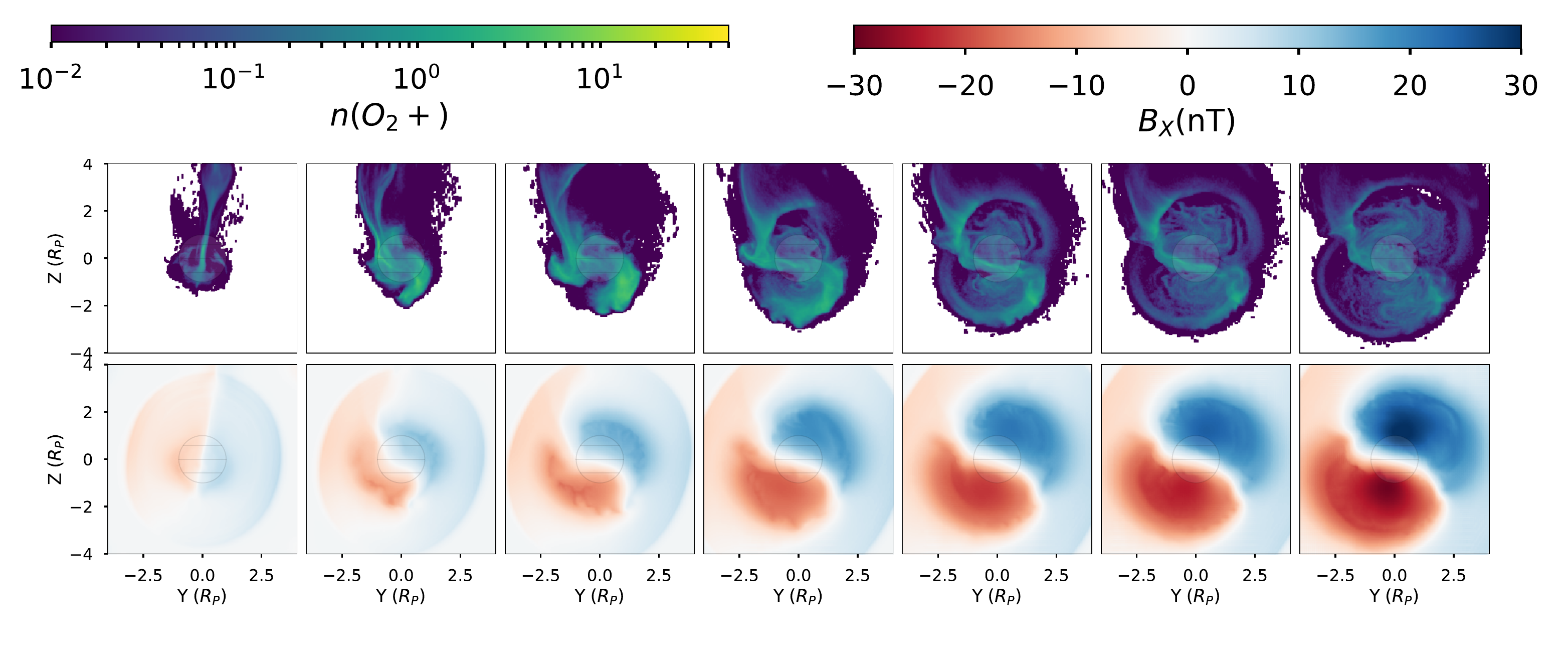}
    \caption{Slices through the tail at $X=-2$ for each simulation (0, 10, 25, 50, 75, 100, 150~nT from left to right). Top row shows $O_2^+$ number density while the bottom row shows the x-component of the magnetic field. The gradual twisting of the tail represents a shift from being IMF draping dominated (symmetric about Y) to dipole dominated (symmetric about Z).}
    \label{fig:s_twist}
\end{figure*}

In this set of simulations the dipole axis and the IMF are roughly perpendicular, making the topology of the current sheet sensitive to the relative strengths of these fields. Figure \ref{fig:s_twist} shows slices through the tail at $x=-2$ for each simulation for both $O_2^+$ number density and the x-component of the magnetic field, which illustrate the location of the current sheet. As the field strength increases the tail becomes more twisted such that the central area is dominated by the dipole contribution while farther out is dominated by the IMF draping. \added{This structure shows the transition in relative influences of the induced and intrinsic magnetospheres, as well as the effect on ion escape morphology.}

\subsection{Magnetic Topology Mapping}
\label{sec:topology_map}

\begin{figure*}
    \centering
    \includegraphics[width=\textwidth]{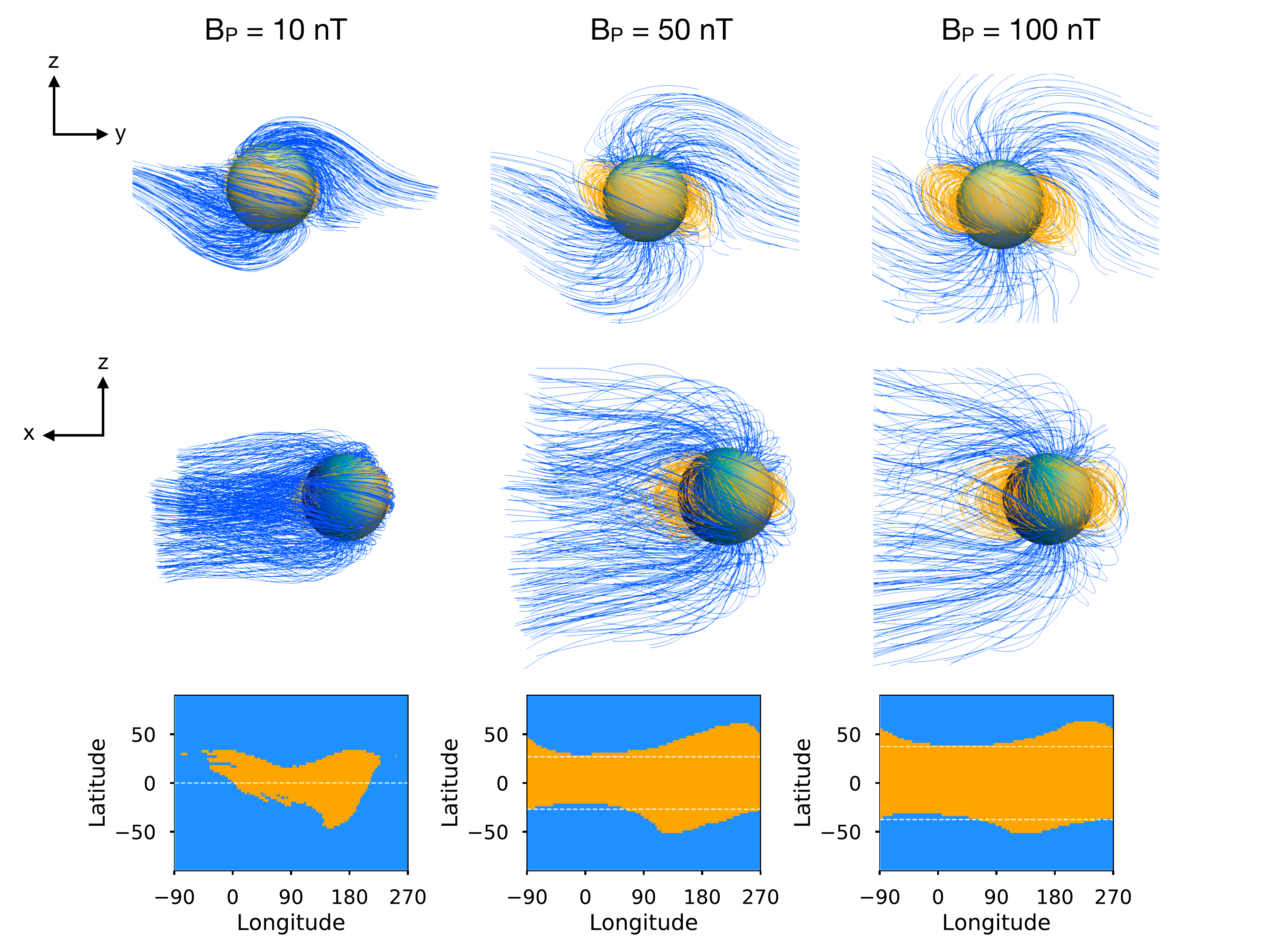}
    \caption{Magnetic field line traces for three representative simulations with planetary magnetic fields of 10~nT (\emph{left}), 50~nT (\emph{center}), and 100~nT (\emph{right}). Top row shows the planet as viewed from the star and middle row shows planet as viewed within the orbital plane; in both cases the motional electric field points upwards. Bottom row is a latitude-longitude map of the field line tracings, with white dashed lines indicating the analytic critical latitude. Closed magnetic field lines are indicated in orange while open magnetic field lines are illustrated in blue. The planet is colored by stellar-zenith angle.}
    \label{fig:topology}
\end{figure*}

Figure \ref{fig:topology} shows magnetic field line tracing through our models with orange lines indicating closed field lines and blue indicating open field lines. The lines were traced in both directions from a spherical shell at 400~km altitude. These field line tracings reflect the above intuition where stronger magnetic fields create larger regions with more closed magnetic field lines.

Because the magnetization of the planet is weak, the interplanetary magnetic field (IMF) carried by the stellar wind has a large impact on the structure of the magnetosphere. This is similar to what has been found at Mercury \citep{SLAVIN20041859, doi:10.1029/97JA03667,KABIN2000397} and Ganymede \citep{doi:10.1029/97GL02201} where the external magnetic field induces polar alignment. Figure \ref{fig:topology} shows this influence where the $+y$ oriented IMF causes the open field lines at the poles to twist towards the dawn ($+y$) and dusk ($-y$) sides in the northern and southern hemispheres respectively, such that at larger distances the field lines align with the IMF. 

The closed field lines attached to the planet on both sides also show twisting to align with the IMF. This causes the closed field lines to attach asymmetrically in longitude; a field line attached at a dawnward longitude in the northern hemisphere will close on the duskward longitude in the southern hemisphere. As the magnetic field strength increases, this effect decreases due to the larger strength of the planetary field in comparison to the IMF.

\added{The lower portion of Figure \ref{fig:topology} shows a latitude-longitude map of the field line topology at the same altitude where the field line tracings began, with dashed lines indicating the critical latitude $\theta_c$ given by Equation \ref{eq:theta_crit}. While field lines outside $\pm\theta_c$ are indicated as open within the analytic formulation described earlier, the field line mapping shows significant area of closed field lines outside this region, particularly for the lowest magnetic field strength (10~nT) and on the night side (longitude $\sim 180$). This occurs because of the aforementioned asymmetries, as well as the neglect of the induced magnetospheric pressure in the analytic formulation.}

\subsection{Magnetic Standoff}
\label{sec:topology}

\begin{figure}
    \centering
    \includegraphics[width=0.45\textwidth]{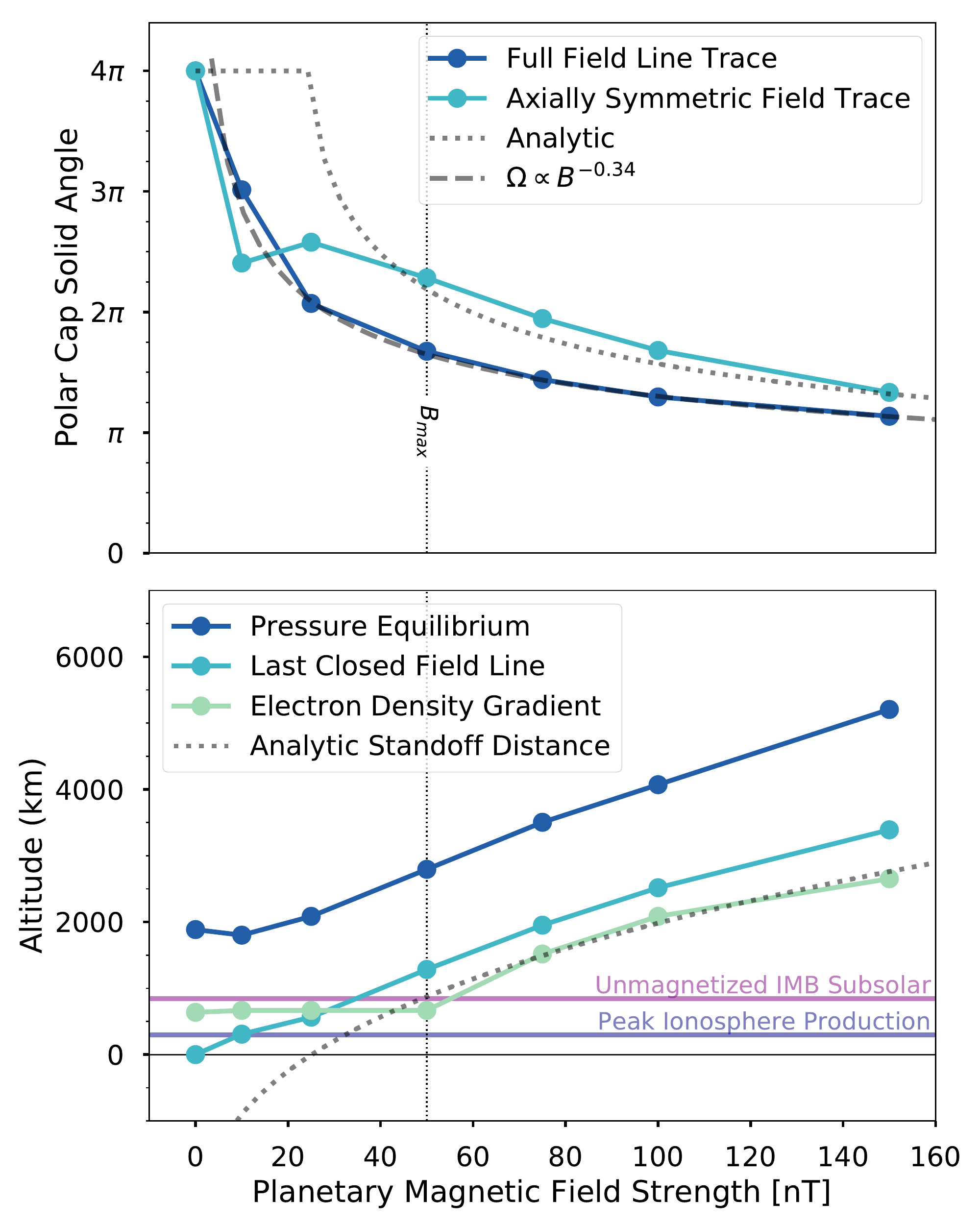}
    \caption{Polar cap solid angle (\emph{top}) and standoff altitude and proxies thereof along the sub-stellar line (\emph{bottom}) plotted over planetary magnetic field. Dotted lines indicate analytic approximations while solid lines with points are calculated from the simulations. Horizontal lines indicate empirical boundaries measured for Mars \citep{2006P&SS...54..357T} and the peak ionosphere production location in the simulations \added{and the vertical line indicates magnetic field associated with peak ion outflow (see Section \ref{sec:ion_esc})}. }
    \label{fig:noon_topology}
\end{figure}

In the upper panel of Figure \ref{fig:noon_topology} the magnetic topology is compared across magnetic field strength by assessing the area of the polar cap open to the stellar wind, an estimate of the region from which ions may escape.  The ``full field line trace'' is calculated by summing the solid angle of all open field lines identified in the tracing associated with Figure \ref{fig:topology}. The ``axially symmetric field trace'' is calculated finding the region of closed field lines along the noon longitude and assuming a polar axis symmetry. The ``analytic cap area" is calculated assuming a polar axially symmetric system with a critical latitude given by Equation \ref{eq:theta_crit}. The discrepancy between the two calculated values is due to the lack of symmetry in the system because of day-night variation, and symmetry breaking induced by the IMF as discussed in the following subsection. Both values are useful to show due the physical accuracy (topology area) and directness of the comparison with the analytic value (symmetric area). 

In general the analytic formulation overestimates the polar cap solid angle, especially at small magnetic field values. This occurs because the analytic estimate must be completely open for all field values where the magnetic field pressure is less than the stellar wind pressure. However, this once again neglects the influence of the induced magnetosphere.

We also show a best fit scaling relationship between the polar cap angle and magnetic field for the full field line trace, where $\Omega_c\propto {B_p}^{-0.34}$. A scaling law exponent of $-1/3$ can be recovered by expanding Equation \ref{eq:cap_angle} around small values of $R_p/R_s$, and then substituting Equation \ref{eq:standoff} for $R_s$. This scaling matches well even to small magnetic field values ($B_P\sim10$~nT).

\added{The bottom panel of Figure \ref{fig:noon_topology} shows the transition from induced to intrinsic magnetospheres through a variety of metrics designed to assess standoff of the stellar wind by the planet. The analytic stand-off altitude (Equation \ref{eq:standoff}) is calculated assuming pressure balance between the incoming stellar wind and the intrinsic magnetosphere of the planet. This is most clearly related to the pressure equilibrium point (the altitude along the sub-stellar line at which the magnetic pressure is equivalent to the stellar wind pressure); however, the pressure equilibrium point occurs at much higher altitude in the magnetosheath due to the expected thickness of the transition region, corresponding to a few proton gyroradii ($\gamma_{H^+}\sim 730$ km) \citep{2004pssp.book.....C}.}

\added{The analytic stand-off altitude is also closely related to the magnetic topology, as shown in Equation \ref{eq:theta_crit}. Here, rather than calculating the latitude associated with the last closed field line, we show the altitude along the sub-stellar line of the last closed field line. This matches fairly well with the analytic stand-off distance but occurs slightly higher due to the the same asymmetry and induced field pressure support as described in Section \ref{sec:topology_map}.}

\added{Finally, we show a measure of the magnetospheric boundary through identification of the steepest gradient in electron number density, a method used to identify the induced magnetospheric boundary at unmagnetized planets \citep{2015GeoRL..42.8885V}. At unmagnetized planets this can be thought of as the ``top'' of the ionosphere, where there is a transition from planetary to stellar wind plasma. At magnetized planets this becomes the outer region of the plasmasphere (see Section \ref{sec:plasmasphere}), where closed magnetic field lines can confine plasma. This measurement thus has a clear relationship to the closed field line boundary, while also having a well defined meaning for unmagnetized planets. }

\added{In Figure \ref{fig:noon_topology} we see that the last closed field line and electron density gradient altitudes agree fairly well with each other and the analytic standoff distance at higher magnetic field strengths. At lower magnetic field strengths, the electron density gradient flattens out due to the intrinsic gradient associated with the ionosphere. This transition occurs where the magnetic stand-off crosses the unmagnetized induced magnetosphere boundary (847~km) \citep{2006P&SS...54..357T}. Critically, this point is also identical to the magnetic field associated with maximum ion outflow as shown in Section \ref{sec:ion_esc}.}

\begin{figure}
    \centering
    \includegraphics[width=0.45\textwidth]{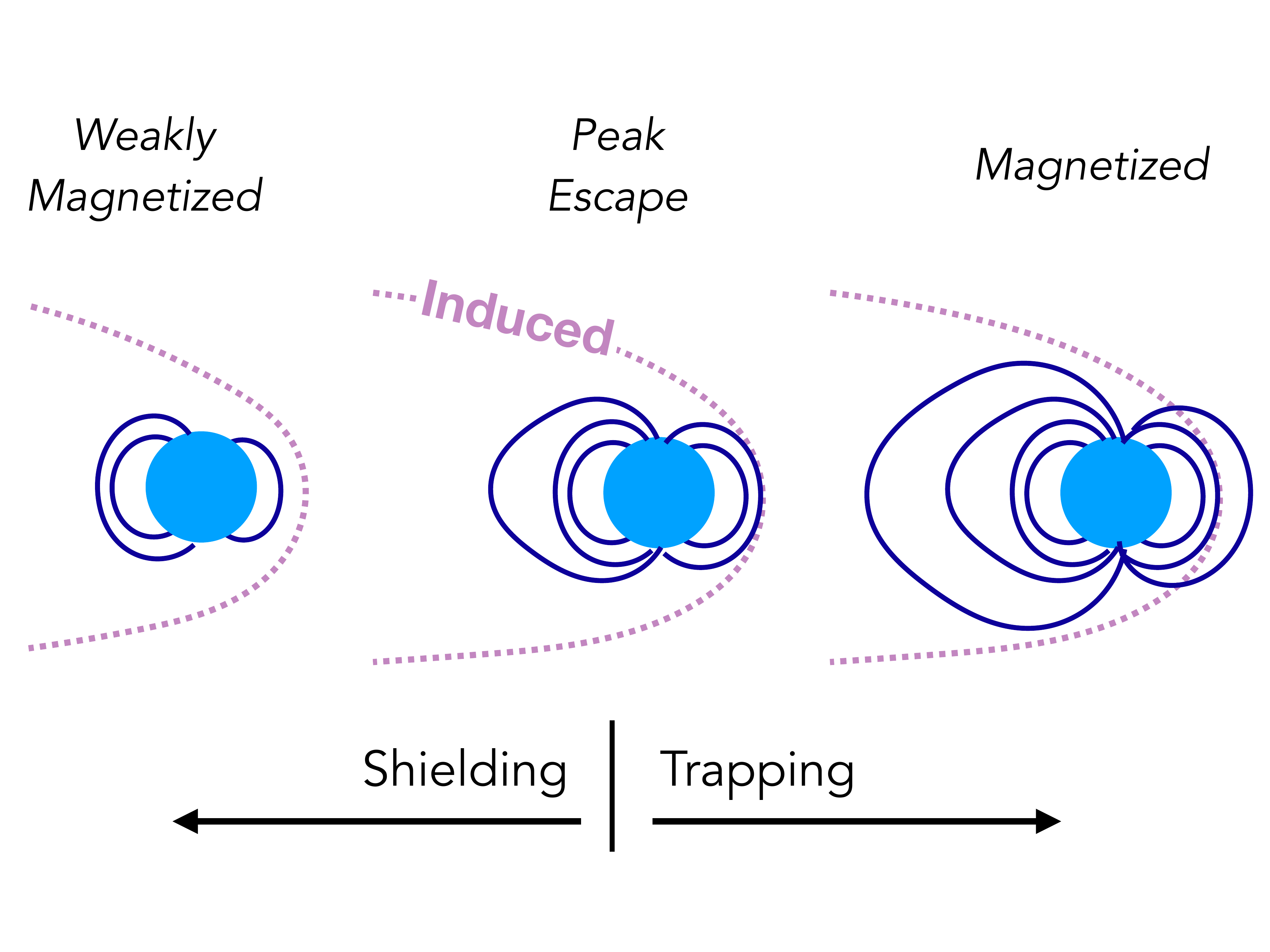}
    \caption{\added{Schematic depiction of the relationship of the induced (\emph{pink dashed}) and intrinsic (\emph{blue solid}) magnetospheric boundaries. The different regimes are responsible for different patterns of ion escape, as discussed in Section \ref{sec:ion_esc}.}}
    \label{fig:boundary_regimes}
\end{figure}

\added{This transition between induced and intrinsic magnetospheres is shown schematically in Figure \ref{fig:boundary_regimes}. The induced magnetosphere boundary remains constant with magnetic field as it is determined by the ionospheric conductivity and pressure support against the stellar wind. For the weakest magnetic fields ($B_P< 50$~nT) the induced magnetosphere is the dominant source of pressure support against the stellar wind, the last closed field lines occur below the boundary, and the electron density gradient is then associated with the induced boundary. At stronger magnetic fields ($B_P > 50$~nT) the intrinsic magnetosphere is the dominant source of pressure support against the stellar wind, the last closed field lines occur above the induced magnetosphere boundary, and the electron pressure gradient is associated with trapped plasma within the closed magnetic field line region. The transition point occurs when the intrinsic magnetosphere standoff distance reaches the induced magnetosphere models; for this set of simulations this occurs at $B_P=50$~nT. In the following section we will discuss the implications this pattern has for ion escape.}

\deleted{The analytic standoff altitude is shown in comparison to the pressure equilibrium point, the last closed field line altitude  (the highest altitude from which a magnetic field line connects to the planet when traced in both directions), and the IMB (the point of steepest gradient in electron number density \citep{2015GeoRL..42.8885V}). Furthermore, two constant boundaries are shown via horizontal lines indicating the peak ionosphere production altitude (300~km) and the average altitude of the IMB for Mars along the sub-solar line (847~km) \citep{2006P&SS...54..357T} for context in an unmagnetized system. While each of these boundaries measures slightly different quantities, they all reflect a transition from planetary plasma to stellar wind plasma, and can thus be used to inform escape rate analysis.}

\deleted{The last closed field line and IMB altitudes agree fairly well with each other and the analytic standoff distance at higher magnetic field strengths. At lower magnetic field strengths the analytic standoff distance neglects the influence of the induced magnetosphere causing it to diverge from the IMB, while the last closed field line is not influenced by the additional pressure support of the IMB.}

\section{Ion Escape}
\label{sec:ion_esc}

As discussed in Section \ref{sec:plasma_env}, the transition between unmagnetized and magnetized planets affects the stand-off of the stellar wind and the opening angle of the polar cap, which in turn affect the pickup ion and cusp escape processes. Here we analyze the trends in ion escape rates and energy and connect them to the influence of the surrounding plasma environment. 

\subsection{Escape Rates and Energies}
\label{sec:esc_rates_energies}
\begin{figure}
    \centering
    \includegraphics[width=0.45\textwidth]{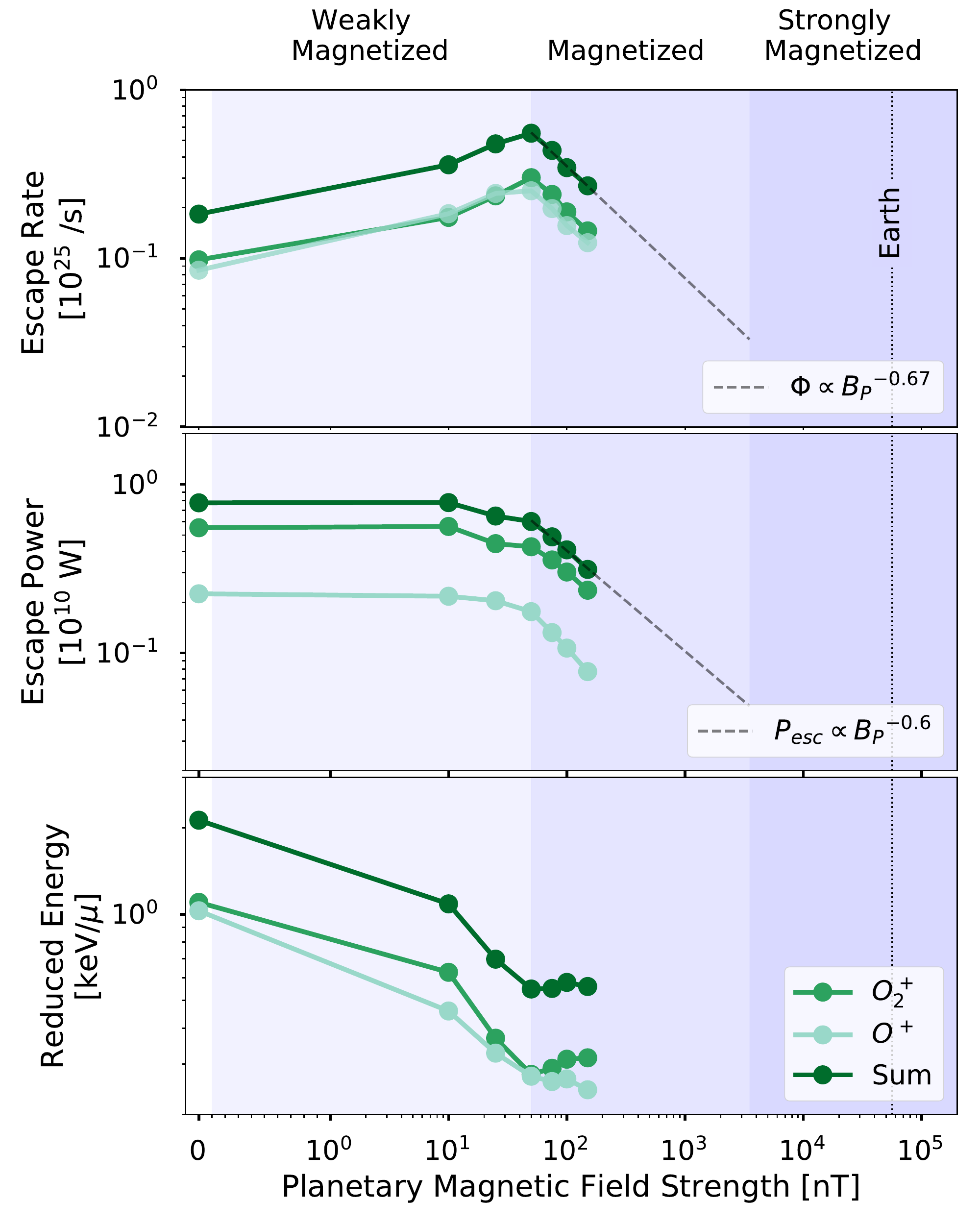}
    \caption{\deleted{Relative} Ion escape rates (\emph{top}), escape power (\emph{middle}), and average ion escape energies (\emph{bottom}) over planetary magnetic field strength for $O_2^+$,$O^+$, and the sum of both ions. Relative escape rates are normalized to the escape rate of the unmagnetized case. Power law fits to the sum are shown in with a dashed line for the escape rate and escape power \added{for $B_p\geq 50~$nT}.}
    \label{fig:rate_compare}
\end{figure}

Figure \ref{fig:rate_compare} shows the escape rate \deleted{relative to the ion flux of the unmagnetized case}, escape power, and average ion escape energy (escape rate divided by escape power) for each simulation run. \added{These parameters are calculated as }

\begin{equation}
    \added{\Phi = \int n_i (\vec{u_i}\cdot\vec{\hat{r}}) r^2 d\Omega}
\end{equation}

\begin{equation}
    \added{P_{esc} =\int m_i n_i {u_i}^2 (\vec{u_i}\cdot\vec{\hat{r}}) r^2 d\Omega}
\end{equation}

\begin{equation}
    \added{\bar{E} = P_{esc}/(\mu \Phi)}
\end{equation}

\added{respectively, where $n_i$ is the ion number density, $u_i$ is the ion velocity, $m_i$ is the ion mass, and $\mu$ is the ion mass expressed in units of the proton mass.} We calculate these rates by integrating over a sphere far from the planet such that the sphere is large enough ($r=4.0R_P$) that the flux value is insensitive to changes in radius, and all outbound flux escapes and does not return to the planet, even for our most strongly magnetized simulation cases. 

\begin{figure*}
    \centering
    \includegraphics[width=\textwidth]{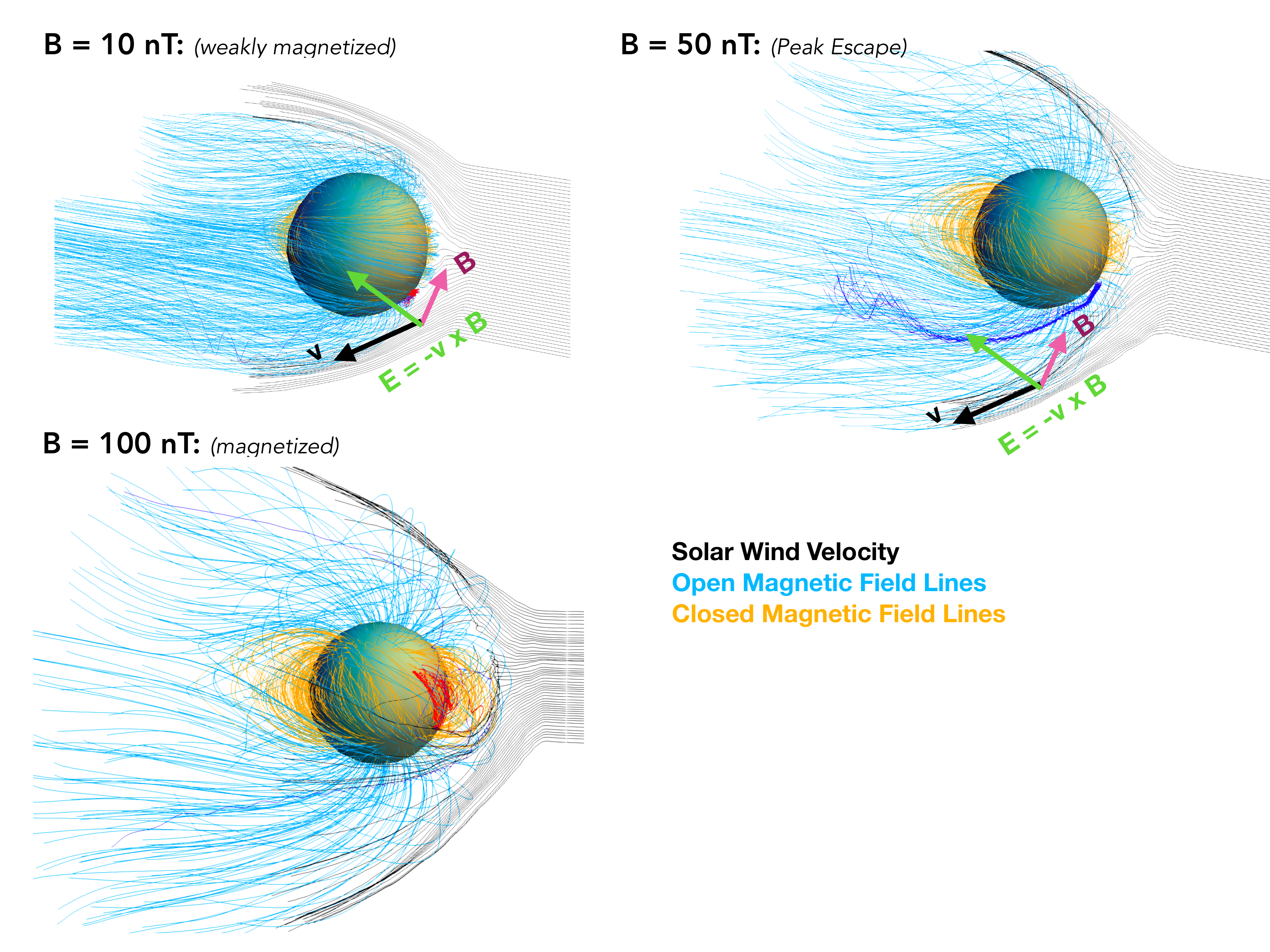}
    \caption{\added{Three example particle trajectory tracings for 10, 50, and 100~nT simulations. Particle tracings are shown in red (trapped) or dark blue (escaping), and magnetic field lines are shown in light blue (open) and orange (closed). While this location was picked primarily as an illustrative example for these three simulations, it is representative of overall ion escape trends discussed in Sections \ref{sec:shielding} and \ref{sec:plasmasphere}.}}
    \label{fig:ptrace}
\end{figure*}

We find that the escape rate increases for magnetic fields up to \replaced{$B_P=50$}{75}~nT and then begins to decrease again, with a maximum variation of a factor of $\sim 3$. \added{The magnetic field associated with peak escape rate ($B_{max}$) is equivalent to the magnetic field where the induced magnetosphere altitude equals the intrinsic magnetosphere altitude. This divides the ion escape trends into different regimes illustrated in Figure \ref{fig:ptrace}; weaker escape associated with predominately induced magnetospheres, stronger escape associated with magnetic shielding and equivalent induced and intrinsic magnetospheres, and weaker escape associated with strong magnetospheres and plasmasphere trapping. Each of these regimes will be explored more fully in the following subsections.} Additionally, we show a power law fit for simulations $B_P\geq 50$~nT where we find that the escape rate scales as ${B_P}^{-0.67}$.

Escape power shows a steady decrease with magnetic field, and a power law scaling of ${B_P}^{-0.60}$ \added{for $B_P\geq 50$~nT}.  Average escape energy is calculated by dividing integrated escape power over the spherical shell by integrated escape flux. \deleted{Reduced escape energy is then calculated by dividing by the ion mass expressed in units of the proton mass ($\mu$).} For both $O_2^+$ and $O^+$ Figure \ref{fig:rate_compare} shows the reduced escape energy is highest for small magnetic field values before reaching an asymptote at around 50~nT. 

\added{Both power laws calculated for escape rate and escape power have clear lower limits at $B_p = 50~nT$, a point we define as $B_{max}$. These power laws also likely have upper limits that may prevent their extension to very strongly magnetized systems like Earth, as indicated by the various shaded regions in Figure \ref{fig:rate_compare}. This is discussed further in Section \ref{sec:scaling_law_limits}.}

\begin{figure}
    \centering
    \includegraphics[width=0.45\textwidth]{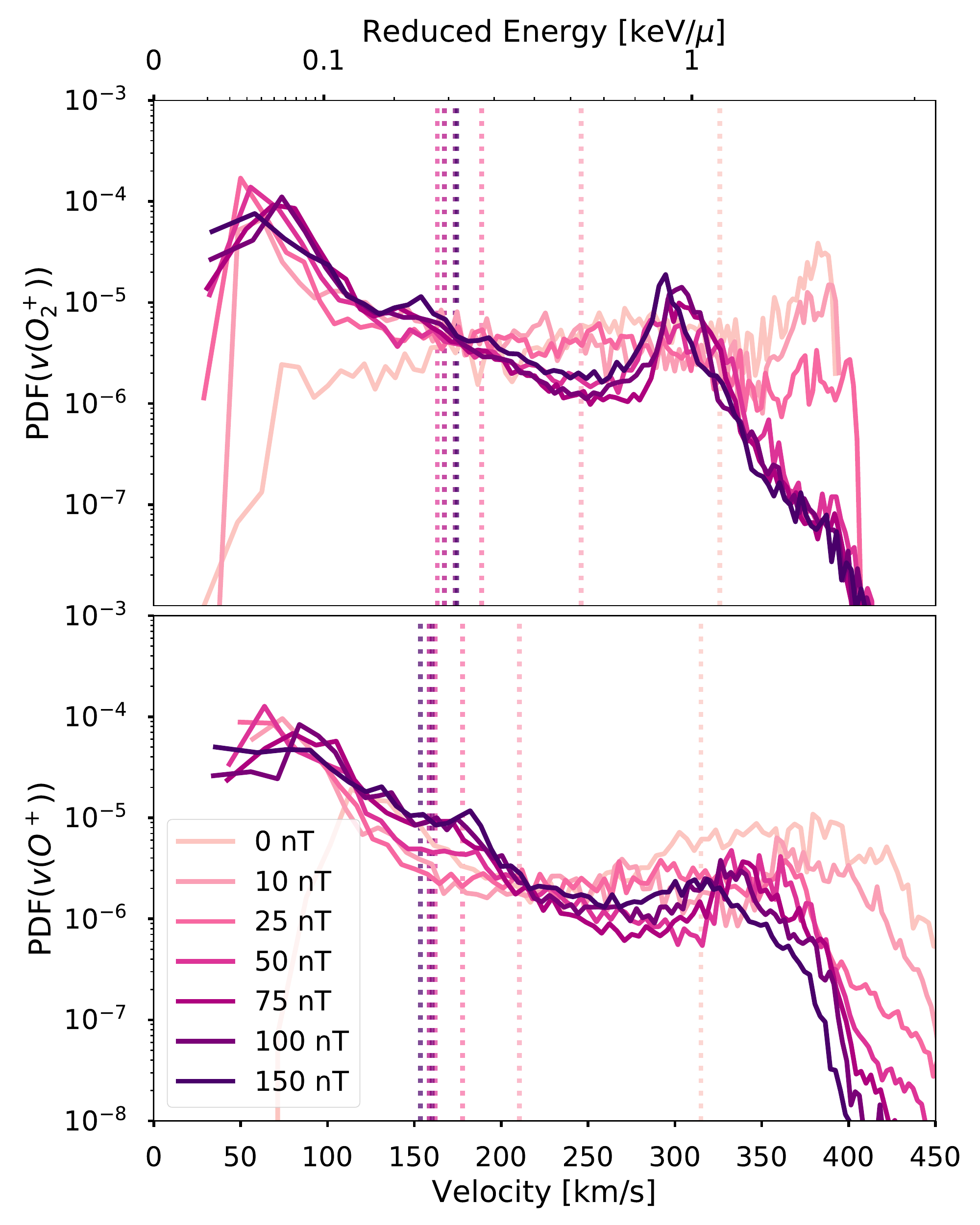}
    \caption{Velocity distribution of escaping $O_2^+$ (\emph{top}) and $O^+$ (\emph{bottom}) ions weighted by number density and velocity, such that integrating over the distribution is equivalent to the escape power. Each color shows a different magnetic field. The dashed lines indicate the weighted average of each distribution (equivalent to the average reduced escape energy plotted in Figure \ref{fig:rate_compare}). }
    \label{fig:velocity_dist}
\end{figure}

Figure \ref{fig:velocity_dist} shows \added{a histogram of the} full velocity distribution of escaping particles through the $R=4.0\;R_P$ sphere, weighted by \replaced{$\textrm{area}\times n_i \times v_i$ and normalized such that the integral over the distribution is 1.}{the number density and velocity such that the integral over the entire distribution will be equivalent to the escape power.} Dashed lines indicate the average escape energies. This figure shows that the decrease in escape energy with increasing magnetic field is caused by a corresponding decrease in high energy particles in the tail of the velocity distributions. As the magnetic field begins to stand off the stellar wind, fewer ions are exposed to the strong motional electric field, and thus fewer ions are accelerated to high velocities.

\subsection{Southern Hemisphere Shielding}
\label{sec:shielding}
The increase in escape rate with increasing magnetic field is caused by stellar wind standoff in the southern hemisphere decreasing the flux of pickup ions accelerated towards the planet and preciptating in the inner boundary. While the motional electric field in the northern hemisphere causes ion pickup and the plume feature, in the southern hemisphere the motional electric field accelerates the ions into the planet, preventing their escape. \added{This effect is illustrated in Figure \ref{fig:ptrace}, with the direction of the motional electric field indicated by the green arrow. While particles injected in the southern hemisphere of the 10~nT simulation are accelerated towards the planet and do not escape, particles injected in the same location in the 50~nT simulation do escape due to the increased stand-off of the stellar wind.}

\begin{figure*}
    \centering
    \includegraphics[width=0.7\textwidth]{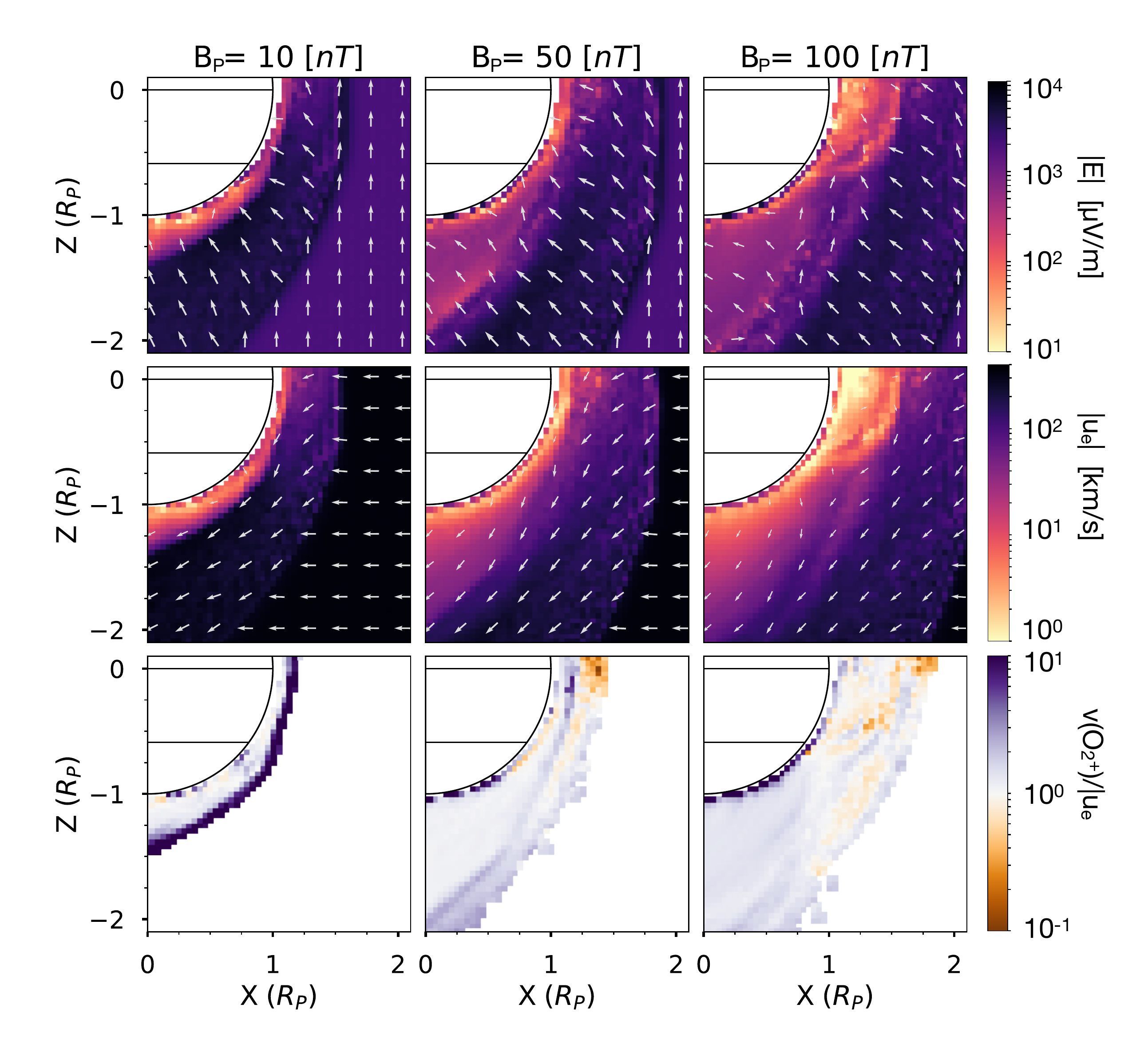}
    \caption{Slices in the $y=0$ plane of motional electric field magnitude (\emph{top}), electron velocity magnitude (\emph{center}), and ratio of the $O_2^+$ velocity to the electron velocity (\emph{bottom}) for the 10~nT (\emph{left}), 50~nT (\emph{center}), and 100~nT (\emph{right}) simulations, showing the effectiveness of the 50 and 100~nT fields in standing off the stellar wind preventing strong planet-oriented electric fields. White arrows indicate direction of the corresponding vector fields.}
    \label{fig:anti_pickup}
\end{figure*}

Figure \ref{fig:anti_pickup} shows slices of the total motional electric field, total electron velocity, and ratio of the fluid velocity and the ion velocity for the 10, 50, and 100~nT simulations. In the 10~nT case strong electric fields due to high electron velocities permeate the region of planetary ions. In the 50 and 100~nT cases the increasing magnetic pressure slows the stellar wind down much more before reaching the planet, decreasing the magnitude of the motional electric field. The ions in these more strongly magnetized cases can then flow farther from the planet along open field lines before being exposed to the strong electric field, and are thus accelerated down-tail rather than into the planet. This effect will saturate roughly when the standoff distance reaches the altitude of the IMB at the terminator, which is illustrated occurring at 50~nT in Figure \ref{fig:noon_topology}. 

\subsection{Plasmasphere Trapping}
\label{sec:plasmasphere}

The decrease in escape rate at higher magnetic field strengths is due to plasmasphere trapping, where ions become trapped within the closed field line regions. \added{This is depicted in Figure \ref{fig:ptrace} where particles escape in the 50~nT simulation, but are confined within the closed magnetic field line in the 100~nT simulation.} This effect is related to the decrease in polar cap solid angle illustrated in the lower panel of Figure \ref{fig:noon_topology}. As more area of the planet is wrapped in stronger closed magnetic field lines, ions will have greater difficulty escaping these closed field regions. 

\begin{figure*}
    \centering
    \includegraphics[width=0.9\textwidth]{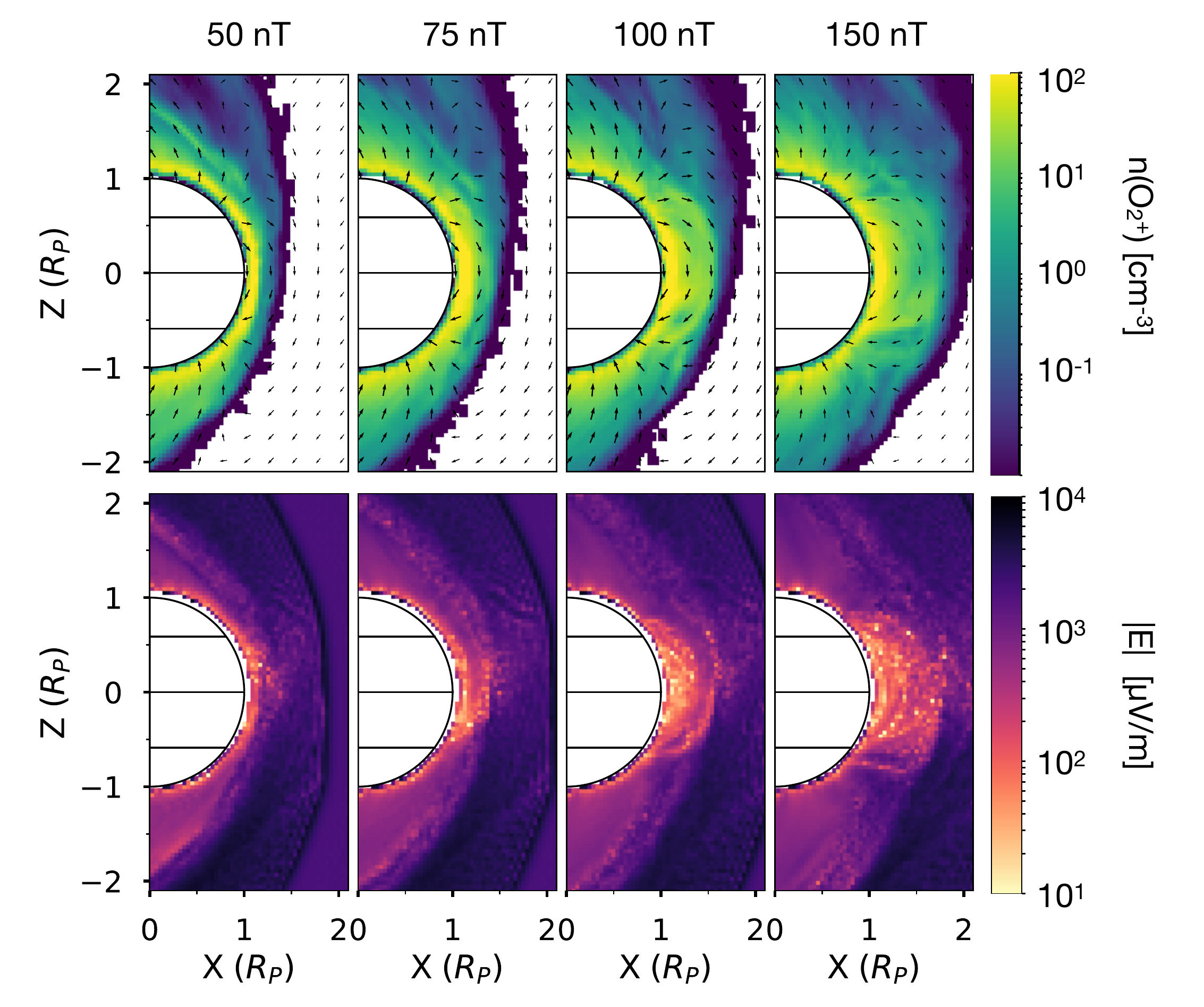}
    \caption{ Slices in the $y=0$ plane showing the development of the plasmasphere. $O_2^+$ number density (\emph{top}) and motional electric field magnitude (\emph{bottom}) for the 50, 75, 100, and 150~nT simulations are shown from left to right, with black arrows over-plotted indicating the direction of the magnetic field.}
    \label{fig:plasmasphere}
\end{figure*}

As the magnetic field of the planet increases creating a region of closed magnetic field lines, the planet develops a region of plasma protected from stellar wind. This toroidal region of dense cold plasma connects to the top of the ionosphere and is known at Earth as the plasmasphere. The development of the plasmasphere is shown in Figure \ref{fig:plasmasphere}, where higher densities of planetary ions occur within the planetary dipole field where the motional electric field becomes small. Because the magnetic field stands off the stellar wind, the motional electric field becomes very small within the the plasmasphere. The ion motions are then dominated by gyromotion along the magnetic field lines, and thus the ions are trapped within the plasmasphere. 

Figure \ref{fig:rate_compare} shows a best fit scaling relation for the escape rate of the plasmasphere trapping dominated simulations ($\geq 50$~nT), where $\Phi\propto {B_P}^{-2/3}$ or $\Phi \propto \Omega_c^2$. This is a steeper relationship than typically expected where $\Phi \propto \Omega_c$ \citep[e.g.][]{gunell_2018}. This is likely due to the production rate variation with stellar zenith angle. Because more ions are produced in the equatorial region, the relationship of escape rate with polar cap angle will be steeper than $\Phi \propto \Omega_c$. Thus, the escape rate relationship with magnetic field will be sensitive to the ion production in relation to the magnetic field topology. This is further discussed in Section \ref{sec:discussion}.








\subsection{Mass Flux and Power Coupling}

Though intrinsic magnetic fields were long thought to shield atmospheres from being stripped by interaction with a stellar wind, it has also been pointed out recently that intrinsic fields increase the cross-sectional area through which a planet interacts with the wind, and therefore the amount of energy or momentum it could ‘collect’ from the wind to drive escape \citep{2013cctp.book..487B}. A common technique used to discuss the relationship between input and output from a given reservoir is through the idea of a coupling efficiency. This idea has also been used in the context of planetary atmospheres where the mass loss rate ($\dot M_{out}$) or the escaping power ($P_{out}$) are coupled to the mass flux ($\dot M_{in}$) or power input ($P_{in}$) from the stellar wind with some efficiency $\epsilon$ \citep[e.g.][]{2005JGRA..110.3221S,2006SSRv..126..209D, doi:10.1002/2017JA024306, egan_2019}. The key questions then become how to define and parameterize the inflow flux/power and identifying the typical efficiency of various processes.  

We put this framework in a more general form by including a correlation constant $k$ such that $\dot M_{out} = \epsilon (\dot M_{in})^k$ (or $P_{out} =  \epsilon (P_{in})^k$), where $k=1$ indicates a traditional linear coupling with a constant efficiency.

We define the inflow properties by scaling the stellar wind parameters with an interaction area defined by the cross-sectional area of the bow shock at the terminator ($x=0$) \added{as} 
\begin{equation}
    \dot M_{in} = A(\rho_{sw}u_{sw})
\end{equation}
\added{and}
\begin{equation}
    \dot P_{in} = A(\rho_{sw}u_{sw}^3)
\end{equation}

\added{where $\rho_{sw}$ and $u_{sw}$ are the density and velocity of the undisturbed stellar wind. The area A} is calculated by identifying the shock boundary in the $x=0$ plane and \replaced{summing the corresponding area in the $x=0$ plane within the boundary.}{ integrating the area contained within the boundary}. \added{We identify the boundary in this way rather than an area calculated using a symmetric shock radius $R_{shock}$ such that $A=\pi R_{shock}^2$ to account for bow shock asymmetry.} As the magnetic field increases the effective obstacle size presented by the planet also increases; this then increases the cross sectional area of the bow shock. The outflow properties are determined in the same method described in earlier in Section \ref{sec:esc_rates_energies}.

\begin{figure*}
    \centering
    \includegraphics[width=\textwidth]{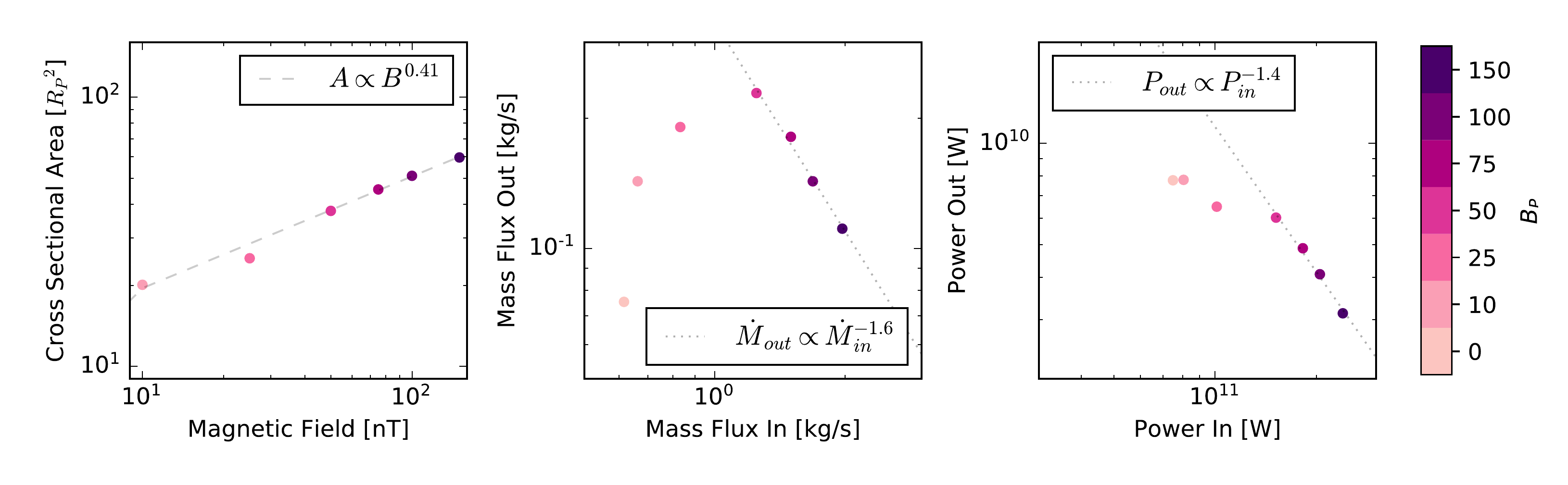}
    \caption{Coupling of the inflow and outflow mass fluxes (\emph{center}) and power (\emph{right}). Outflow properties are calculated over a spherical shell far from the planet, while inflow properties are calculated assuming a constant inflow from the stellar wind over an interaction region determined by the cross-sectional area at the terminator (\emph{left}). Dashed lines indicate a coupling of the form $\dot M_{out} = \epsilon (\dot M_{in})^k$ (or $P_{out} =  \epsilon (P_{in})^k$).}
    \label{fig:power_coupling}
\end{figure*}

The resulting inflow and outflow fluxes and power for each simulation are shown in Figure \ref{fig:power_coupling}, as well as the cross sectional areas calculated for each simulation. We find that the cross sectional area increases with magnetic field as roughly $A\propto {B_P}^{0.41}$ or $A\propto R_S^{1.23}$ (by Equation \ref{eq:standoff}). 

Neither the mass flux nor the power are coupled with a linear efficiency, nor does either relationship show a constant power law correlation over the entire range. This is indicative of the same change in ion escape processes as described in the preceding sections on escape trends, despite a constant scaling in cross-sectional area. For the higher magnetic field values where plasmasphere trapping is the dominant process affecting escape, the mass flux and power correlations have indexes of $k=-1.60$ \added{($\epsilon=-0.48$)} and $k=-1.43$ \added{($\epsilon=25.8$)} respectively. As the inflow mass (power) scales with area ($A\propto {B_P}^{0.41}$) and the outflow mass (power) scales as $\propto {B_P}^{-0.66}$ ($\propto {B_P}^{-0.6}$), both correlation constants follow straightforwardly.

\section{Discussion}
\label{sec:discussion}

\subsection{Stellar Driving Dependence}
\label{sec:disc_stellar_driving}
All the simulations shown in this paper were run with the same nominal Martian solar wind driving; however, stellar winds are dynamic and can dramatically change ion loss \citep{doi:10.1002/2017JA024306,2017EGUGA..1911139B}. Here we discuss the impact of our choice in stellar wind on our results. 

\replaced{As shown in Equation \ref{eq:standoff}, the stellar wind pressure controls the standoff distance associated with an intrinsic magnetosphere. The induced magnetosphere boundary has also been shown to depend weakly on dynamic pressure and extreme ultra-violet input, but much less so than the intrinsic magnetosphere \citep{doi:10.1002/2017JA024098}. As the magnetic field associated with the peak escape rate, $B_{max}$, occurs when the intrinsic magnetosphere reaches the induced magnetosphere, it will also scale with stellar wind pressure as $B_{max}\propto P_{sw}^{1/2}$. }{The stellar wind pressure controls the magnetic standoff distance and thus the polar cap area as shown in Equations \ref{eq:standoff} and \ref{eq:theta_crit}. As both the southern hemisphere shielding and plasmasphere trapping effects are dependent on the standoff and polar cap area, the magnetic field strength associated with peak escape will also vary with the stellar wind pressure as $B_{max}\propto P_{sw}^{1/2}$. } Therefore a planet experiencing a stellar wind pressure increase \replaced{by orders of magnitude}{of an order of magnitude or two}, as may be likely for a planet in the habitable zone around an M-dwarf \citep{2015MNRAS.449.4117V,2017ApJ...843L..33G},\replaced{ will have a much larger $B_{max}$.}{ although the increased power input of the planet must also be taken into account along with production or diffusion limitations for interpreting the total escape rates for such a system \citep{egan_2019}.}

As shown in Section \ref{sec:plasma_env}, due to the relative strengths of the planetary dipole and the IMF, the IMF critically affects the magnetic topology of the system. The IMF can be described by a magnitude, the clock angle $\phi$ (angle in the YZ plane, perpendicular to the sub-stellar line), and the cone angle $\theta$ (angle between the IMF and the sub-stellar line). Our study has assumed an IMF that is perpendicular to the dipole axis, thus a motional electric field that is aligned with the dipole axis. In the opposite case where the IMF is aligned with the dipole axis, the southern hemisphere shielding effect would be mitigated as the motional electric field will stand-off or accelerate into the plasmasphere rather than the poles. Thus, there is a critical dependence on the IMF clock angle relative to the dipole axis, necessitating further study for different configurations.

Similarly, the angle between the dipole magnetic field axis and the sub-stellar line also impacts the escape rate relations we have shown here. The ion production method we have chosen is dependent on the stellar zenith angle, such that the highest production occurs at the sub-stellar point. A dramatically different configuration would occur for a planetary pole aligned with this peak in ion production; however, this is a less likely scenario due to spin-orbit alignment. Further, planets orbiting close to their stars have fast orbital velocities. In the case when the orbital velocity is the same as the stellar wind speed, the upstream stellar wind vector and the sub-stellar line have an angle of 45$^\circ$.

\subsection{Model Limitations}
\label{sec:disc_ion_limits}
In a global hybrid model like RHybrid, specific escape rates depend on the inner boundary ionospheric emission rates. The emission rate of inospheric ions is a free parameter, which is fixed by comparing a model to in situ observations from solar system planets \citep[e.g.][]{2009AnGeo..27.4333J,2018JGRA..123.1678J}. \replaced{While we show escape rates here that are comparable to current escape rates at Mars \citep{lundin1990aspera, 2017JGRA..122.9552J,2015GeoRL..42.9142B}, we caution against the over-interpretation of these rates due to their dependence on the lower boundary, and instead concentrate on the relative differences.}{Therefore, the preceding analysis has concentrated on comparing relative ion escape rates. }

Furthermore, our implementation of ionospheric emission is driven by a predefined Chapman ion production profile. Currently, accurately resolving ionospheric dynamics on the same domain as the global magnetosphere is computationally infeasible. Although some simulations suites include a one way coupling of an ionosphere implementation to a global magnetosphere \citep[e.g.][]{2009JGRA..114.5216G,2016JGRA..12110190B,2016JGRA..121.6378M}, we chose to restrict our ionospheric production to a simple approximation. This allows us to focus on the magnetospheric physics associated with an increasing magnetic field, without introducing an additional layer of uncertainty from coupling an additional model.

We have also neglected to include an ionospheric electrodynamics model, and instead use a constant resistivity value above the lower boundary and zero resistivity at the lower boundary. Global thermospheric structure  \citep{1982JGR....87.1599R} and ionospheric-magnetospheric coupling \citep{2004AnGeo..22..567R} are both affected by ionospheric electrodynamics through mechanisms such as current closure, atmospheric Joule heating, and Alfven wave dissipation. 

Impact ionization, neutral structure, and resistivity are all affected by auroral precipitation  \citep{1995JGR...10017153C,1980JGR....85.2185D, 1979JGR....84.1855T, 1987JGR....92.2565R, 1987JGR....92.7606F}, which will change with magnetic field strength. In the weakly magnetized cases we have chosen here the polar cap solid angle (and the corresponding precipitation per area) varies by a factor of a few across the simulations. While this is likely to become important at large magnetic field strengths (see Section \ref{sec:disc_ion_limits}), it is likely a secondary effect behind the basic ion production method we use here.

\subsection{Comparison With Existing Results}

In general our conclusions agree with similar results from previous analytic and numerical studies, but add a deeper understanding of the physics controlling the transition from unmagnetized to weakly magnetized planets. 

An analytic study by \citet{gunell_2018} showed a similar peak in ion escape rates at the point where the standoff radius reaches the IMB; however, their model enforces a sharp transition in escape processes at this point rather than a gradual build up. This study also uses the typical assumption of linear dependence of ion escape with the polar cap angle, whereas we find a stronger dependence.

\citet{2012EP&S...64..149K} used a hybrid model to study weakly magnetized planets and found a peak ion escape rate for a Mars-like planet with a 10~nT magnetic field. Our peak escape rate occurs at a higher point due to the stronger stellar wind conditions. Furthermore, the same group studied the magnetic morphology of the same system and found similar trends in tail twisting and IMF control; however, the effects were weaker due to the relative weakness of the IMF in comparison with the dipole strengths for a similar stellar wind pressure \citep{2008P&SS...56..823K}.

\citet{2018MNRAS.481.5146B} studied the impact of magnetic fields on stellar wind erosion of planetary atmospheres using an analytic estimation of the maximum escape rate for a given system. This model assumes there is a direct relationship between the inbound and outbound mass flux, an assumption our results contradict in the weakly magnetized regime; however, they largely concentrate on magnetic fields estimated for ancient Earth, which are larger than those studied here.

Other studies examining the escape rates of exoplanets with and without magnetic fields have been performed \citep[e.g.][]{2017ApJ...837L..26D}; however, these results are not directly comparable as the magnetic fields studied are much stronger than the range considered here.

\subsection{Limits of Derived Scaling Laws}
\label{sec:scaling_law_limits}

\begin{table*}
\begin{tabular}{@{}clll@{}}
\toprule
Parameter & Scaling & Lower Limit      & Upper Limit                                                                                                                                                                                                         \\ \midrule
$\Omega$  & $\propto B^{-0.33}$             & $R_s<R_P$        & N/A                                                                                                                                                                                                                   \\
$\dot M$  & $\propto  B^{-0.67}$             & $R_s\sim R_{IMB}$, $B\sim B_{max}$ &  \begin{tabular}[c]{@{}l@{}}a) may flatten beyond $P_B \sim 35 P_{sw}$\\ b) cusp dynamics may dominate past $P_B \sim (5.26)^6 P_{sw}$\end{tabular} \\
$P$       & $\propto  B^{-0.60}$             & $R_s\sim R_{IMB}$, $B\sim B_{max}$ &                                                                                                                                                                                                              a), b)       \\
$A$       & $\propto  B^{0.41}$              & $R_s\sim R_{IMB}$, $B\sim B_{max}$ & Bow shock models needed                                                                                                                                                                                                                   \\ \bottomrule
\end{tabular}

\caption{Summary of derived scaling relationships for polar cap solid angle ($\Omega$), mass outflow ($\dot M$), escape power ($P_{esc}$), and bow shock interaction area ($A$) with magnetic field along with upper and lower limits of the scaling law. \added{Lower limits are calculated empirically from the simulations while upper limits are postulated as in the text.}}
\label{tab:scaling_summary}
\end{table*}

In Table \ref{tab:scaling_summary} we summarize the scaling laws we have fit for various parameters from the simulations with $B_P\geq 50$~nT, along with estimations for the lower and upper limits of the power law applicability. Although we can calculate error bars on the scaling relations from the model results, these errors do not reflect the sensitivity of the calculated power law to the model assumptions. Instead, here we discuss some of the main parameters and assumptions these laws are most sensitive to.

As shown in Section \ref{sec:topology}, the polar cap solid angle should scale as ${B_P}^{-1/3}$ for large magnetic field values. This matches our calculated scaling law very well, even to small magnetic field values leading us to place an approximate lower limit of $R_S\sim R_P$. While magnetic topology is a fairly straightforward feature to interpret from our models, it is also sensitive to the magnetic dipole orientation with respect to the IMF and sub-stellar line. This relationship does not have a clear upper limit for reasonable values of planetary magnetic field; however, as the polar cap region becomes smaller care must be taken to understand the asymetries and dependence on IMF \citep[e.g.][]{2004JGRA..109.5222K, doi:10.1029/2007GL029357}

We find a mass loss rate that scales more strongly with magnetic field than predicted, likely due to the variation in ion production with SZA. This result is quite sensitive to the ion production mechanism, as well as the magnetic topology. From the simulations here we have shown the lower limit will be $R_S\simeq R_{IMB}$ \added{or equivalently $B_P \simeq B_{max}$}. 

The upper limit is more difficult to estimate. First, the power law will become more shallow with decreasing polar cap angle, likely approaching the predicted $\dot M \propto \Omega \propto B^{-1/3}$ relationship. Additionally, as the polar cap angle becomes small, cusp dynamics associated with ionospheric electrodynamics and energization from particle precipitation will become more dominant. \citet{2005JGRA..110.3221S} showed that cusp ionospheric outflow at the Earth is associated with polar cap Poynting flux and electron precipitation leading to Joule dissipation and electron heating, causing increased ion upwelling and subsequent outflow. Both of these effects are associated with a large power input to a small polar cap area coupling to the ionosphere. 

A polar cap solid angle an order of magnitude smaller than our base case will occur at  $\Omega_c\sim 0.1\times 4\pi$, or $P_B\sim (5.26)^6 P_{SW}$. For this set of simulations this would occur for magnetic fields of $B_P\sim 3500$~nT, which is still much smaller than the Earth's surface field, where cusp escape processes are known to be relevant \citep{2012GeoRL..3918102L}. While the escaping ion flux may in fact continue to decrease past this point, the dominant physical mechanisms are likely to be quite different due to the intensity of precipitating particle flux in the small auroral region and the scale of the electrodynamic systems.

As shown in Figure \ref{fig:rate_compare}, the average escape energy per particle becomes constant with increasing magnetic field and the corresponding total power of ion outflow decreases as $P_{esc}\propto B^{0.6}$. Thus, if the average energy of escaping particles remains constant, the $P_{esc}$ relationship is determined by the escape flux relationship, and will have the same upper and lower limits. As the escape energy is determined by the stellar wind velocity and corresponding motional electric field, this is likely to be true as long as the wind maintains its momentum through an encounter with the planetary obstacle. This will be true in the absence of significant mass-loading.

The interaction area is found to scale as $A\propto B^{0.41}$, which is shallower than one might expect for an interaction area that scales with a length scale ($R_S$) squared, or $A\propto B^{2/3}$. While we recover the standoff radius relationship as expected (see Figure \ref{fig:noon_topology}), the interaction area is more sensitive to the stellar wind parameters and corresponding shock properties. Furthermore, the shape of the shock will be affected by the shape of the obstacle; the assumption of a spherically symmetric obstacle is particularly poor for small magnetic field values (as shown in Section \ref{sec:plasma_env}).

\section{Conclusions}
\label{sec:conclusions}

Understanding the influence of planetary magnetic field on ion escape is a critical area for assessing the potential habitability of exoplanets. Here we have presented a systematic study of the influence of a weak magnetic dipole on the plasma environment and corresponding ion escape from a Mars-like exoplanet using a hybrid plasma model.

We have found that an intrinsic magnetic field does not always act to decrease the ion escape rate, and instead leads to an increase in escape rate up until the point where the magnetic standoff reaches the terminator IMB before then beginning to decrease again. This reflects a balance between southern hemisphere shielding of ions from the motional electric field and increased plasmasphere trapping due to smaller polar cap area. The value of the magnetic field associated with maximum ion escape is therefore dependent on the stellar wind pressure.

Where possible we have calculated scaling laws for the relationship of various fundamental quantities and magnetic field. The scaling laws are interpreted in the global context of planetary atmospheric escape theory. Lower limits for these scaling laws have been derived from the simulation results, while upper limits are postulated via predicted limits on the observed processes. 

Going forward, more study is needed to assess effects associated with the dynamics of the stellar wind for planets with weak magnetospheres due to the extreme influence of the stellar wind on these systems, as well as a larger array of IMF configurations relative to the planetary dipole. Furthermore, the effects of stronger planetary magnetic fields must be assessed using a simulation platform that fully couples a polar wind model. 

\section{Acknowledgments}
Hilary Egan was supported by the Department of Energy Computational Science Graduate Fellowship, Maven , and a generous donation from Marcia Grand. This research used resources of the National Energy Research Scientific Computing Center (NERSC), a U.S. Department of Energy Office of Science User Facility operated under Contract No. DE-AC02-05CH11231. Global hybrid simulations were performed using the RHybrid simulation platform, which is available under an open source license by the Finnish Meteorological Institute (https://github.com/fmihpc/rhybrid).

\bibliographystyle{plainnat}
\bibliography{bfield}
\end{document}